\documentclass{llncs}
\usepackage{xspace}
\usepackage{longtable}
\usepackage{pgfplots}
\usepackage{amsmath}
\usepackage{paralist}
\usepackage{times}
\usepackage[hyphens,obeyspaces,spaces]{url} %
\usepackage{multirow}
\usepackage{rotating}

\newcommand{\nop}[1]{}

\newcommand{\etal}{et~al.\xspace}

\newcommand{\ccfont}[1]{\ensuremath{\mathsf{#1}}}
\newcommand{\ccfontx}[1]{\ccfont{#1}\xspace}

\newcommand{\NP}[0]{\ccfontx{NP}}
\newcommand{\SigmaI}[1]{\ccfontx{\Sigma_{#1}P}}
\newcommand{\PSPACE}[0]{\ccfontx{PSPACE}}

\begin{document}
\title{Conformant Planning as a Case Study of \\ Incremental QBF Solving\thanks{Supported by the Austrian Science Fund (FWF) under
grants S11409-N23 and P25518-N23 and the German Research Foundation (DFG) under grant ER 738/2-1. This article appeared in the \textbf{proceedings} of
\emph{Artificial Intelligence and Symbolic Computation (AISC)}~\cite{DBLP:conf/aisc/EglyKLP14}. An \textbf{extended version} appeared in \emph{Annals of Mathematics
and Artificial Intelligence}~\cite{Egly2016}.}}
\titlerunning{Conformant Planning as a Case Study of Incremental QBF Solving}  %
\author{Uwe Egly\inst{1} \and Martin Kronegger\inst{1} \and Florian Lonsing\inst{1} \and Andreas Pfandler\inst{1,2}}
\authorrunning{Egly et al.} %
\tocauthor{Uwe Egly, Martin Kronegger, Florian Lonsing, and Andreas Pfandler}
\institute{Institute of Information Systems, Vienna University of Technology, Austria\\
\and
School of Economic Disciplines, University of Siegen, Germany \\
\email{firstname.lastname@tuwien.ac.at}
}

\maketitle              %

\begin{abstract}
We consider planning with uncertainty in the initial state as a case study of
\emph{incremental} quantified Boolean formula (QBF) solving. We report on experiments
with a workflow to incrementally encode a planning instance into a
sequence of QBFs. To solve this sequence of successively constructed QBFs, we
use our \emph{general-purpose incremental} QBF solver DepQBF. Since the
generated QBFs have many clauses and variables in common, our approach avoids  
redundancy both in the encoding phase and in the solving phase. Experimental
results show that incremental QBF solving outperforms non-incremental QBF
solving. Our results are the first empirical study of  
\emph{incremental} QBF solving in the context of planning and motivate its use in 
  other application domains.
\end{abstract}
\section{Introduction}
\label{sect:introduction}

Many workflows in formal verification and model checking rely on
certain logics as languages to model verification conditions or
properties of the systems under consideration. Examples are
propositional logic (SAT), quantified Boolean formulas (QBFs), and
decidable fragments of first order logic in terms of satisfiability
modulo theories (SMT).  A tight integration of decision procedures to
solve formulas in these logics is crucial for the overall performance
of the workflows in practice.

In the context of SAT, \emph{incremental solving}~\cite{DBLP:conf/sat/AudemardLS13,DBLP:journals/entcs/EenS03,DBLP:conf/sat/LagniezB13,DBLP:conf/sat/NadelRS14} has become a state of the
art
approach. 
Given a sequence of related propositional formulas $S =
\langle\phi_0,\phi_1,\ldots,\phi_n\rangle$ an incremental SAT solver reuses information that was gathered
when solving $\phi_i$ in order to solve the next formula $\phi_{i+1}$. Since 
incremental solving avoids some redundancy in the process of solving the sequence
$S$, it is desirable to integrate incremental solvers in practical
workflows. In contrast, in \emph{non-incremental solving} the solver does not keep any
information from previously solved formulas and always starts from scratch.

QBFs allow for explicit universal ($\forall$) and
existential ($\exists$) quantification over Boolean variables. The problem of checking the satisfiability of QBFs is \PSPACE-complete. We consider QBFs as
a natural modelling language for planning problems with uncertainty in
the initial state. 
In \emph{conformant planning} we are given a set of state variables over a specified domain, a set of actions with preconditions and effects, an initial state where some values of the variables may be unknown, and a specification of the goal.
The task is to find a sequence of actions, i.e., a plan, that leads from the initial state to a state where the goal is satisfied.
Many natural problems, such as repair and therapy planning~\cite{SmithW98}, can be formulated as conformant planning problems.
When restricted to plans of length polynomial in the input size this form of planning is \SigmaI{2}-complete~\cite{DBLP:journals/ai/BaralKT00}, whereas classical planning is \NP-complete.
Therefore, using a transformation to QBFs in the case of conformant planning is a very natural approach.
Rintanen~\cite{DBLP:conf/aaai/Rintanen07} presented such transformations.
Recently, Kronegger~\etal~\cite{KroneggerPP13} showed that transforming the planning instance into a sequence of QBFs can be competitive.
In this approach, they generated a QBF for every plan length under consideration and invoked an external QBF solver on each generated QBF. 
However, the major drawback is that the QBF solver cannot reuse information from previous runs and thus has to relearn all necessary information in order to solve the QBF.
In this work we overcome this problem by tightly integrating a 
\emph{general-purpose incremental QBF solver} in an incremental workflow to solve planning problems.
To obtain a better picture of the performance gain through the incremental approach, we perform a case study where we compare incremental and non-incremental QBF solving on benchmarks for conformant planning.\smallskip

\noindent The \textbf{main contributions} of this work are as follows.
\begin{compactitem}
\item \emph{Planning tool.} We present a planning tool based on the transformation of planning instances with unknown variables in the initial state to QBFs. 
This tool implements an incremental and exact approach, i.e., it is guaranteed to find a plan whenever a plan exists and -- if successful -- it returns a plan of \emph{minimal} length. 
Furthermore, our tool allows for the use of arbitrary (incremental) QBF
solvers. 
\item \emph{Experimental evaluation.} We evaluate the performance of the incremental and the non-incremental approach to planning with incomplete information in the initial state.
Thereby, we rely on incremental and non-incremental variants of the QBF solver DepQBF~\cite{DBLP:conf/cp/LonsingE14,DBLP:conf/icms/LonsingE14}.\footnote{DepQBF is free software:
    \url{http://lonsing.github.io/depqbf/}} 
Incremental QBF solving outperforms non-incremental QBF solving in our planning tool. 
Our results are a case study of incremental QBF solving and motivate its use in other application domains.
In addition, we also compare our results to heuristic approaches.
\end{compactitem}

\section{Incremental QBF Solving}
\label{sect:incremental:QBF:solving}

We focus on QBFs $\psi = \hat{Q}.\phi$ in prenex conjunctive normal form
(PCNF). All quantifiers occur in the prefix $\hat{Q} = Q_1B_1 \ldots Q_nB_n$
and the CNF part $\phi$ is a quantifier-free propositional formula in CNF. The prefix
consists of pairwise disjoint sets $B_i$ of quantified Boolean variables, where $Q_i \in
\{\forall,\exists\}$, and gives rise to a linear ordering of the variables: we
define $x < y$ if $x \in B_i$, $y \in B_j$ and $i < j$.

The semantics of QBFs is defined recursively based on the quantifier types and
the prefix ordering of the variables. The
QBF consisting only of the truth constant \emph{true} ($\top$) or \emph{false}
($\bot$) is satisfiable or unsatisfiable, respectively. The QBF $\psi =
\forall x.\,\psi'$ with the universal quantification $\forall x$ at the
leftmost position in the prefix is
satisfiable if $\psi[x := \bot]$ and $\psi[x := \top]$ are satisfiable, where
the formula $\psi[x := \bot]$ ($\psi[x := \top]$) results from $\psi$ by
replacing the free variable $x$ by $\bot$ ($\top$). The QBF $\psi = \exists x.\,\psi'$ with the
existential quantification $\exists x$ is satisfiable if $\psi[x :=
\bot]$ or $\psi[x := \top]$ are satisfiable.

Search-based QBF solving~\cite{DBLP:journals/jar/CadoliSGG02} is a
generalization of the DPLL algorithm~\cite{DPLL} for SAT. Modern search-based
QBF solvers implement a QBF-specific variant of conflict-driven clause
learning (CDCL) for SAT, called
QCDCL~\cite{DBLP:journals/jair/GiunchigliaNT06,DBLP:conf/tableaux/Letz02,DBLP:conf/sat/LonsingEG13,DBLP:conf/cp/ZhangM02}. In
QCDCL the variables are successively assigned until an (un)satisfiable subcase
is encountered. The subcase is analyzed and 
  new learned constraints (clauses or cubes) are inferred by 
Q-resolution~\cite{DBLP:journals/iandc/BuningKF95,DBLP:conf/sat/LonsingEG13}. The purpose of the learned
constraints is to prune the search space and to speed up proof search. Assignments are retracted by
backtracking and the next subcase is determined until
the formula is solved. 

Let $\langle \psi_0,\psi_1,\ldots,\psi_n \rangle$ be a sequence of QBFs. In
\emph{incremental QBF solving based on QCDCL}, we must keep track
which of the constraints that were learned on a solved QBF $\psi_i$ can be
reused for solving the QBFs $\psi_j$ with $i<j$.  An approach to
incremental QBF solving was first presented in the context of bounded model checking~\cite{DBLP:conf/date/MarinMLB12}. We rely on the
general-purpose incremental QBF solver DepQBF~\cite{DBLP:conf/cp/LonsingE14,DBLP:conf/icms/LonsingE14}.

To illustrate the potential of incremental QBF solving, we present a case study of QBF-based conformant planning in the following sections. To this end we discuss conformant planning and two types of benchmarks used in the experimental analysis.

\section{Conformant Planning and Benchmark Domains}
\label{sect:planning}

A conformant planning problem consists of a set of state variables over a specified domain, a set of actions with preconditions and effects, an initial state where some values of the variables may be unknown, and a specification of the goal.
The task is to find a sequence of actions, i.e., a plan, that leads from the initial state to a state where the goal is satisfied.
The plan has to reach the goal for all possible values of unknown variables,
i.e., it has to be fail-safe. 
This problem can nicely be encoded into QBFs, e.g., by building upon the encodings by Rintanen~\cite{DBLP:conf/aaai/Rintanen07}.
Conformant planning naturally arises, e.g., in repair and therapy
planning~\cite{SmithW98}, where a plan needs to succeed even if some obstacles~arise.

The length of a plan is the number of actions in the plan.
As one is usually looking for short plans, the following strategy is used.
Starting at a lower bound $k$ on the minimal plan length, we iteratively increment the plan length $k$ until a plan is found or a limit on the plan length is reached.
This strategy is readily supported by an incremental QBF solver because a large number of clauses remains untouched when moving from length $k$ to $k+1$ and always leads to \emph{optimal} plans with respect to the plan length.

The two benchmark types we consider in our case study are called ``Dungeon''.
These benchmarks are inspired by adventure computer-games and were first presented at the 
QBF workshop 2013~\cite{KroneggerPP13}. 
In this setting a player wants to defeat monsters living in a dungeon.
Each monster requires a certain configuration of items to be defeated.
In the beginning, the player picks at most one item from each pool of items.
In addition, the player can exchange several items for one more powerful item if she holds all necessary ``ingredients''.
Eventually, the player enters the dungeon.
When entering the dungeon, the player is forced to pick additional items.
The dilemma is that the player does not know which items she will get, i.e., the
additional items are represented by variables with \emph{unknown} values in the
initial state. 
It might also happen that the new items turn out to be obstructive given the previously chosen item configuration.
The goal is to pick items such that irrespective of the additional items she
defeats at least one monster.

We consider two variants of the Dungeon benchmark. 
In variant \emph{v0} the player is only allowed to enter the dungeon once, thus has to pick the items and build more powerful items in advance. %
In contrast, in variant \emph{v1} the player might attempt fighting the monsters several times and pick/build further items in between if she was unsuccessful.

Despite the simple concept, these benchmarks are well suited for our case study.
First, they capture the full hardness of \SigmaI{2}-complete problems.
Second, it is natural to reinterpret the game setting as a configuration or maintenance problem.

\section{QBF Planning Tool}
\label{sect:planning:tool}

We briefly describe our planning tool that takes planning instances as input
and encodes them as a sequence of QBFs.
This tool generates a plan of minimal length for a given conformant planning instance with uncertainty in the initial state.

\begin{figure}[t]%
\centering
\includegraphics[width=1\linewidth]{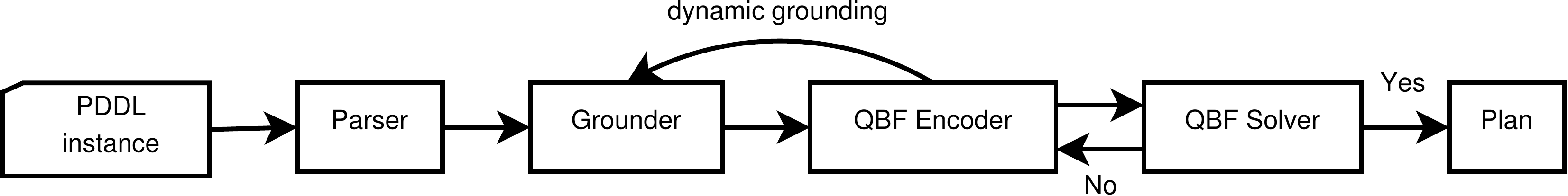}
\caption[Architecture or our solver]{Architecture of the planning tool.}
\label{fig:architectureOfSolver}
\end{figure}

Figure~\ref{fig:architectureOfSolver} illustrates the architecture of our planning tool which was used for the experiments. 
The tool takes a planning instances given in PDDL format as input.
After parsing the input, the grounder analyzes the given planning instance and calculates a lower bound $\ell$ on the plan length. 
Starting with a plan length of $k=\ell$, the grounder then grounds only relevant parts of the instance, i.e., the grounder systematically eliminates variables from the PDDL instance. 
In a next step, the QBF encoder takes the ground representation as input and transforms it into a QBF that is satisfiable if and only if the planning problem has a plan of length $k$. 
The encoding which is used for this transformation to QBFs builds upon the
$\exists\forall\exists$-encoding described in the work of
Rintanen~\cite{DBLP:conf/aaai/Rintanen07}.
We decided to employ the $\exists\forall\exists$-encoding rather than a $\exists\forall$-encoding as this gives a more natural encoding and simplifies the PCNF transformation.
Since in this work we focus on a comparison of the incremental and non-incremental approach, we do not go into the details of the encoding.
After the transformation into a QBF, the QBF encoder then invokes a QBF solver on the generated QBF.
If the generated QBF is satisfiable, our system extracts the optimal plan %
from the assignment of the leftmost $\exists$-block.
If the QBF is unsatisfiable, the plan length $k$ is incremented, additional relevant parts of the problem may need grounding, and the subsequent QBF is passed to the solver. 
Below, we give an overview of the features and optimizations of our planning tool.

Since grounding the planning instance can cause an exponential blow-up in the size of the input, we have implemented a dynamic grounding algorithm.
This algorithm uses ideas from the concept of the planning graph~\cite{DBLP:journals/ai/BlumF97} to only ground actions that are relevant for a certain plan length.
With this optimization, we are able to make the grounding process feasible. 
Although the planning tool provides several methods to compute lower bounds on
the plan length, in our experiments we always started with plan length $0$ to
allow for a better comparison of the incremental and non-incremental approach.

Our \emph{incremental} QBF solver DepQBF is written in C whereas the planning tool is
written in Java. To integrate DepQBF in our tool and to employ its features
for incremental solving, we implemented a Java interface for DepQBF, called
\emph{DepQBF4J}.\footnote{DepQBF4J is part of the release of DepQBF version
  3.03 or later.} This way, DepQBF can be integrated into arbitrary Java
applications and its API functions  
 can then be called via the Java Native Interface
(JNI). 

In our planning 
tool, the use of DepQBF's API is crucial for incremental solving because we
have to avoid writing the generated QBFs to a file. Instead, we add and modify
the QBFs to be solved directly via the API of DepQBF. The API provides
\emph{push} and \emph{pop} functions to add and remove \emph{frames}, 
i.e., sets of clauses, in a stack-based manner. The CNF part of a QBF is
represented as a sequence of frames.

Given a planning instance, the workflow starts with plan length $k = 0$.  The
QBF $\psi_k$ for plan length $k$ can be encoded naturally in an incremental
fashion by maintaining two frames $f_0$  and $f_1$ of clauses: clauses which encode the goal state are
added to $f_1$. All other clauses are added to $f_0$. Frame $f_0$ is
added to DepQBF before frame $f_1$. 
If $\psi_k$ is unsatisfiable, then $f_1$ is deleted by a \emph{pop}
operation, i.e., the clauses encoding the goal state of plan length $k$ are removed.
The plan length is increased by one and additional clauses
encoding the possible state transitions from plan length $k$ to $k+1$ are
added to $f_0$.  The clauses encoding the goal state for plan length
$k+1$ are added to a new $f_1$. 
Note that in the workflow clauses are added to $f_0$ but this frame is never deleted.

The workflow terminates if
\begin{inparaenum}[(1)]
\item \label{workflow-termination-1}  the QBF $\psi_k$ is satisfiable, indicating
that the instance has a plan with \emph{optimal} 
length $k$, or 
\item \label{workflow-termination-2} $\psi_k$ is unsatisfiable and $k+1$ exceeds a user-defined upper
bound, indicating that the instance does not have a plan of length
$k$ or smaller, or 
\item \label{workflow-termination-3} the time or memory limits are exceeded. 
\end{inparaenum}
In the cases
(\ref{workflow-termination-1}) and (\ref{workflow-termination-2}), we consider the planning instance as solved. For the experimental
evaluation, we imposed an upper bound of 200 on the plan length.

The Dungeon benchmark
captures the full hardness of problems on the second level of the polynomial
hierarchy. Therefore, as shown in the following section, already instances with moderate plan lengths might be hard for QBF solvers as well as for 
planning-specific solvers~\cite{KroneggerPP13}. We considered an upper bound
of 200 of the plan length to be sufficient to show the difference between the
incremental and non-incremental QBF-based approach.  The 
hardness is due to the highly combinatorial nature of the Dungeon instances, which 
 also applies to configuration and maintenance problems.  Further, 
configuration and maintenance problems can be encoded easily into conformant planning 
 as the Dungeon benchmark is essentially a configuration
problem.

Our planning tool can also be combined with any \emph{non-incremental} QBF
solver to determine a plan of minimal length in a \emph{non-incremental}
fashion.  This is done by writing the QBFs which correspond to the plan lengths $k
= 0,1,\ldots$ under consideration to separate files and solving them with a
standalone QBF solver~\cite{KroneggerPP13}.

\section{Experimental Evaluation}
\label{sect:experiments:incqbf}

We evaluate the incremental workflow described in the previous section using
planning instances from the Dungeon benchmark. The purpose of our
experimental analysis is to compare incremental and non-incremental QBF
solving in the context of conformant planning. Thereby, we provide the first
empirical study of \emph{incremental} QBF solving in the planning domain. 
 In addition
to~\cite{DBLP:conf/sat/MarinMB12,DBLP:conf/date/MarinMLB12}, our results
independently motivate  the use of incremental QBF solving in other application domains.

From the Dungeon benchmark described in Section~\ref{sect:planning}, we selected 144 planning instances from each
variant \emph{v0} and \emph{v1}, resulting in 288 planning instances. Given a
planning instance, we allowed 900 seconds wall clock time and 7 GB of memory
for the entire workflow, which includes grounding, QBF encoding and QBF solving. All experiments reported were run on AMD Opteron 6238, 2.6 GHz, 64-bit
Linux.

\begin{table}[t]
\centering
\begin{tabular}{|r@{\quad}c@{\quad}c@{\quad}c@{\quad}c@{\quad}c@{\quad}c@{\quad}c|}
\hline
\multicolumn{8}{|c|}{\emph{288 Planning Instances (Dungeon Benchmark: \emph{v0} and \emph{v1})}} \\
 & Time & Solved & Plan found & No plan &
 $\overline{t}$ & $\overline{b}$ & $\overline{a}$ \\
DepQBF: & 112,117 & 168 & 163 & \ \,5 & 24.40 & 2210 & 501,706 \\
incDepQBF: & 103,378  & 176 & 163 & 13 & 14.55 & \ \,965 & 120,166 \\
\hline
\end{tabular}
\caption{Overall statistics for the planning workflows implementing
  incremental and non-incremental QBF solving by incDepQBF and DepQBF,
respectively: total time for the workflow on all 288 instances (including time
outs), solved instances, solved instances where a plan was found and where no plan
with length 200 or shorter exists, average time ($\overline{t}$) in seconds, number of
backtracks ($\overline{b}$) and
assignments ($\overline{a}$) performed by the QBF solver on the solved
instances. 
}
\label{tab:compare:inc:noninc:dungeons:overview}
\end{table}
\begin{table}[t]
\centering
\begin{tabular}{|r@{\quad}c@{\quad}c@{\quad}c@{\quad}c@{\quad}c@{\quad}c|}
\hline
\multicolumn{7}{|c|}{\emph{Uniquely Solved Planning Instances}} \\
 & Solved & Plan found & No plan &
 $\overline{t}$ & $\overline{b}$ & $\overline{a}$ \\
DepQBF: & \ \,2 & 0 & \ \,2 & 545.04 & \ \,99 & 1,024,200 \\
incDepQBF:  & 10 & 0 & 10 &  \ \,94.15 & 174 & \ \ \ \ \,45,180 \\
\hline
\end{tabular}
\caption{Statistics like in
  Table~\ref{tab:compare:inc:noninc:dungeons:overview} but for those planning
  instances which were uniquely solved when using either incDepQBF or
  DepQBF, respectively. For all of these uniquely solved instances, no plan
was found within the given upper bound of 200.}
\label{tab:compare:inc:noninc:dungeons:unique}
\end{table}
\begin{figure}[t]
\includegraphics[scale=0.5]{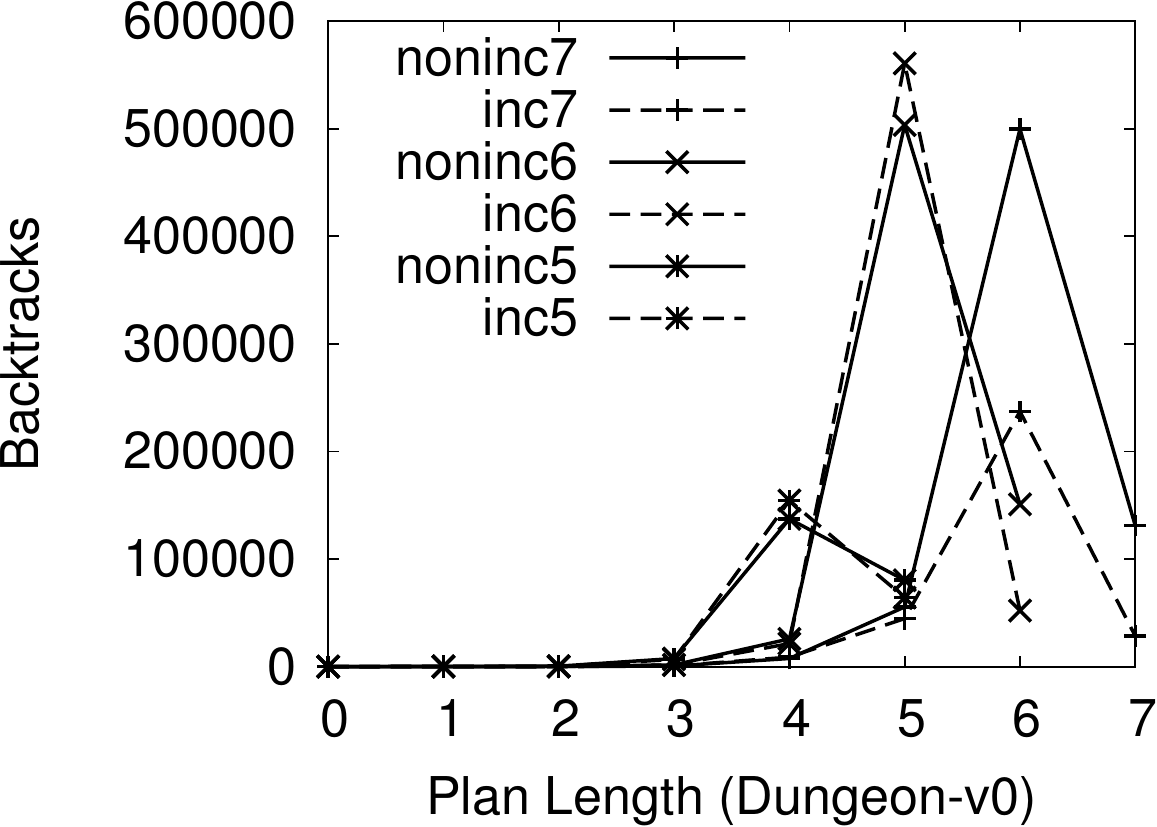}
\hfill
\includegraphics[scale=0.5]{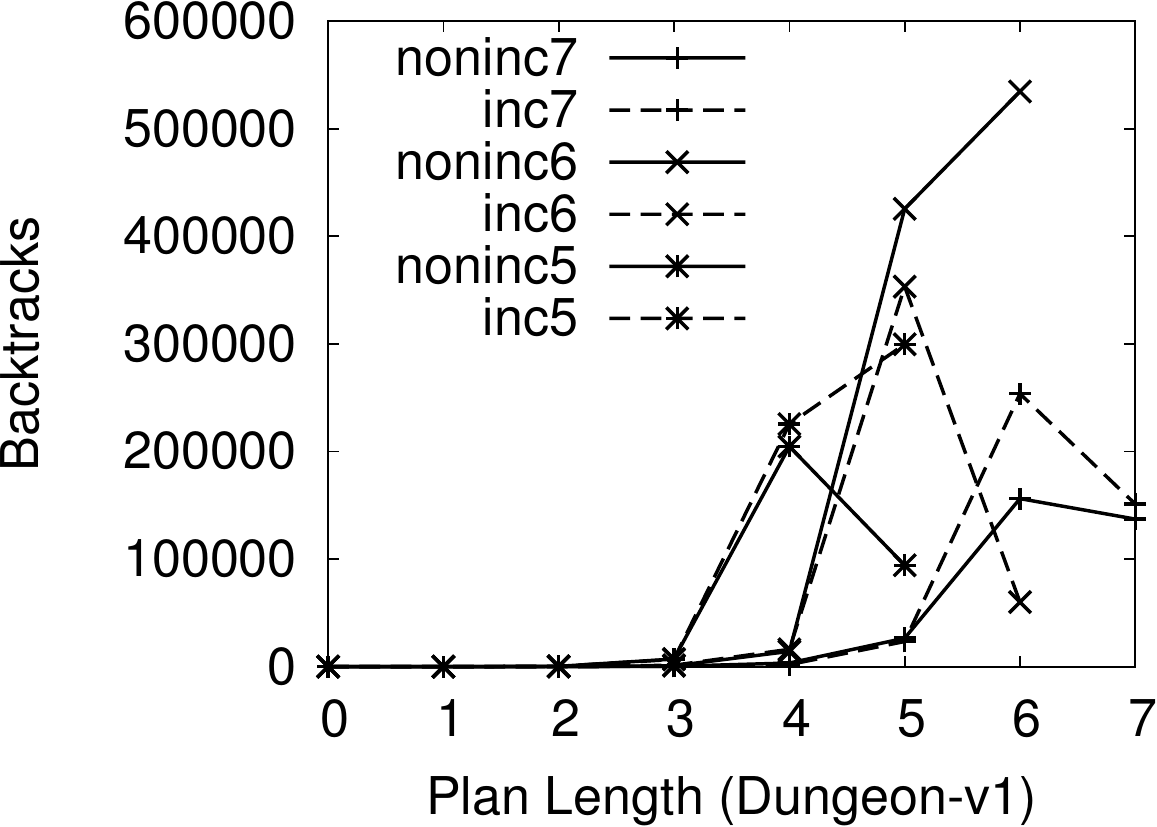}
\caption{Related to 
Table~\ref{tab:compare:inc:noninc:dungeons}. Let $P_5$, $P_6$ and $P_7$ be the sets of planning instances where a plan of length 5, 6, and 7
  was found using both incDepQBF and DepQBF. The data
  points on the lines ``inc5'' (dashed)  and ``noninc5'' (solid)  show the total numbers of
  backtracks spent by incDepQBF and DepQBF on the QBFs corresponding to
  the plan lengths $i = 0,\ldots,5$ for all instances in $P_5$. The data
  points for $P_6$ and $P_7$ were computed similarly for the plan lengths $i =
  0,\ldots,6$ and $i = 0,\ldots,7$, respectively, and are shown on the lines
  ``inc6'', ``noninc6'' and ``inc7'', ``noninc7''. }
\label{fig:histogram:inc:noninc}
\end{figure}

We first compare the performance of incremental and non-incremental QBF
solving in the planning workflow. To this end, we used incremental and non-incremental
variants of our QBF solver DepQBF, referred to as incDepQBF and
DepQBF, respectively.  For non-incremental solving, we called the
standalone solver DepQBF by system calls from our planning
tool. Thereby, we generated the QBF encoding of a particular planning
instance and wrote it to a file on the hard disk. DepQBF then reads the QBF from
the file. For incremental solving, we called incDepQBF through its API via the
DepQBF4J interface. This way, the QBF encoding is directly added to incDepQBF
by its API within the planning tool (as outlined in the previous section), and no files are written. 
The solvers incDepQBF and DepQBF have the \emph{same}
codebase. Therefore, differences in their performance are due to whether
incremental solving is applied or not. 

\begin{figure}[t]
\begin{center}
  \begin{minipage}{0.49\textwidth}
    \includegraphics[scale=0.5]{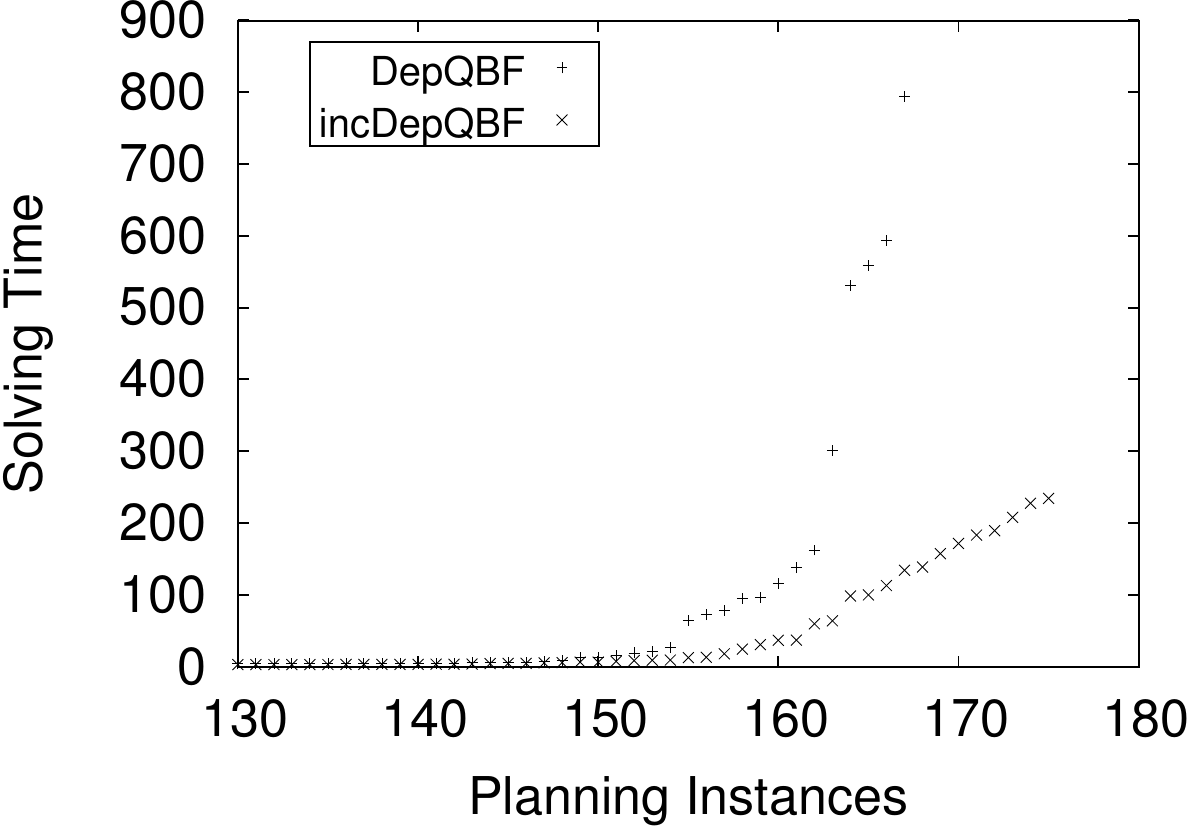}
    \caption{Sorted run times of the workflow with 
      incremental (incDepQBF) and non-incremental solving (DepQBF), 
      related to Table~\ref{tab:compare:inc:noninc:dungeons:overview}.}
    \label{fig:times:inc:noninc:full}
  \end{minipage}
\hfill
  \begin{minipage}{0.49\textwidth}
    \includegraphics[scale=0.5]{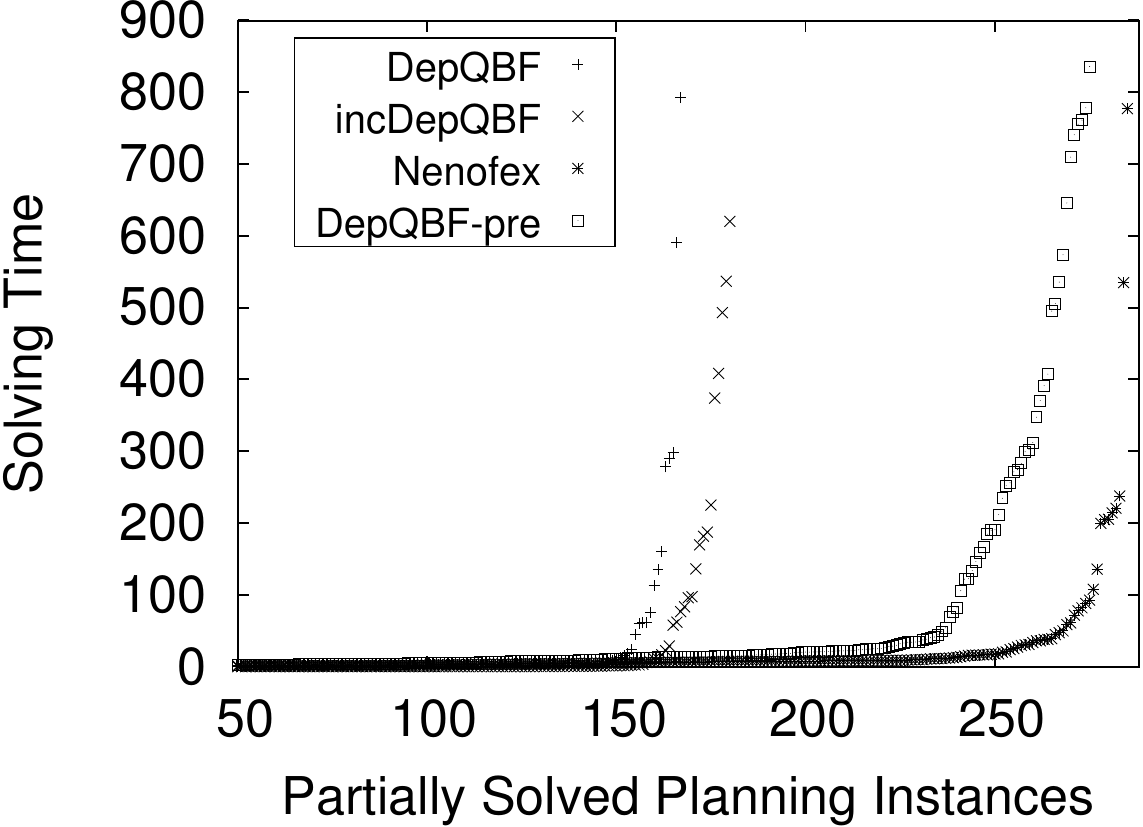}
    \caption{Sorted accumulated run times of solvers on selected QBFs
      from the planning workflow. DepQBF-pre includes preprocessing.}
    \label{fig:times:inc:noninc:standalone}
  \end{minipage}
\end{center}
\end{figure}

The statistics in
Tables~\ref{tab:compare:inc:noninc:dungeons:overview} to~\ref{tab:compare:inc:noninc:dungeons} 
and Figure~\ref{fig:times:inc:noninc:full} illustrate
that incremental QBF solving by incDepQBF outperforms 
non-incremental solving by DepQBF in the  workflow in terms of solved instances, 
 uniquely solved instances (Table~\ref{tab:compare:inc:noninc:dungeons:unique}), 
run time and in the number of backtracks and assignments spent in QBF solving.
With incDepQBF and DepQBF, 166
instances were solved by both. For three of these 166 instances, no plan exists.

The different calling principles of incDepQBF (by the API) and DepQBF (by
system calls) may have some influence on the overall run time of the workflow,
depending on the underlying hardware and operating system. In general, the use
of the API avoids I/O overhead in terms of hard disk accesses and thus might
save run time. Due to the timeout of 900 seconds and the relatively small
number of QBF solver calls in the workflow (at most 201, for plan length 0 up
to the upper bound of 200), we expect that the influence of the calling
principle on the overall time statistics in
Tables~\ref{tab:compare:inc:noninc:dungeons:overview}
and~\ref{tab:compare:inc:noninc:dungeons:unique} and
Figure~\ref{fig:times:inc:noninc:full} is only marginal.  Moreover, considering
backtracks and assignments as shown in
Table~\ref{tab:compare:inc:noninc:dungeons} as an independent measure of the
performance of the workflow, incremental solving by incDepQBF clearly
outperforms non-incremental solving by DepQBF.

Figure~\ref{fig:histogram:inc:noninc} shows how the number of backtracks
 evolves if the plan length is
increased. On the selected instances which have a plan with optimal length
$k$, we observed peaks in the number of backtracks by incDepQBF and
DepQBF on those QBFs which correspond to the plan length $k-1$. Thus empirically the final unsatisfiable QBF for plan length $k-1$ 
 is harder to solve than the QBF for the optimal plan length $k$ or
shorter plan lengths. Figure~\ref{fig:histogram:inc:noninc} (right) shows
notable exceptions. For $k = 6$, the number of backtracks by DepQBF increases
in contrast to incDepQBF. For $k = 5$ and $k = 7$, incDepQBF spent more
backtracks than DepQBF. We attribute this difference to the heuristics in
(inc)DepQBF. The same QBFs must be solved by incDepQBF and DepQBF in one run
of the workflow. However, the heuristics in incDepQBF might be negatively
influenced by previously solved QBFs. We
made similar observations on instances not solved with either incDepQBF or
DepQBF where DepQBF reached a longer plan length than \mbox{incDepQBF} within the
time limit.
\begin{table}[t]
\begin{minipage}{0.5\textwidth}
\centering
\begin{tabular}{|ccccc|}
\hline
\multicolumn{5}{|c|}{\emph{Dungeon-\emph{v0} (81 solved instances)}} \\
& & DepQBF & incDepQBF & \ diff. (\%) \\
\hline
\multirow{3}{*}{\begin{sideways}\emph{Total}\end{sideways}} & $a$: & 171,245,867 & 122,233,046 & -28.6  \\
& $b$: & 1,660,296 & 1,237,384 & -25.4  \\
& $t$: & 1,253.50 & 638.94 & -49.0   \\
\hline
\multirow{6}{*}{\begin{sideways}\emph{Per instance}\end{sideways}} & $\overline{a}$: & 2,114,146 & 1,509,049 & -28.6  \\
& $\overline{b}$: & 20,497 & 15,276 & -25.4  \\
& $\overline{t}$: & 15.47 & 7.88 & -49.0   \\
\cline{3-5}
& $\tilde{a}$: & 1,388 & 1,391 & +0.2   \\
& $\tilde{b}$: & 13 & 11 & -15.3   \\
& $\tilde{t}$: & 1.01 & 0.37 & -63.8   \\
\hline
\multirow{6}{*}{\begin{sideways}\emph{Per solved QBF}\end{sideways}} & $\overline{a}$: & 629,580 & 449,386 & -28.6  \\
& $\overline{b}$: & 6,104 & 4,549 & -25.4  \\
& $\overline{t}$: & 4.60 & 2.34 & -49.0  \\
\cline{3-5}
& $\tilde{a}$: & 828 & 833 & +0.6  \\
& $\tilde{b}$: & 1 & 1 & +0.0  \\
& $\tilde{t}$: & 1.01 & 0.36 & -63.8  \\
\hline
\end{tabular}
\end{minipage}
\begin{minipage}{0.5\textwidth}
\centering
\begin{tabular}{|ccccc|}
\hline
\multicolumn{5}{|c|}{\emph{Dungeon-\emph{v1} (82 solved instances)}} \\
& & DepQBF & incDepQBF & \ diff. (\%) \\
\hline
\multirow{3}{*}{\begin{sideways}\emph{Total}\end{sideways}} & $a$: & 183,674,291
& 164,131,257 & -10.6 \\
& $b$: & 1,670,375 & 1,459,655 & -12.6 \\
& $t$: & 1,308.26 & 773.39 & -40.8 \\
\hline
\multirow{6}{*}{\begin{sideways}\emph{Per instance}\end{sideways}} &
$\overline{a}$: & 2,239,930 & 2,001,600 & -10.6 \\
& $\overline{b}$: & 20,370 & 17,800 & -12.6 \\
& $\overline{t}$: & 15.95 & 9.43 & -40.8 \\ 
\cline{3-5}
 & $\tilde{a}$: & 1,595 & 1,641 & +2.8 \\
& $\tilde{b}$: & 15 & 15 & +0.0 \\
& $\tilde{t}$: & 1.31 & 0.37 & -71.7 \\
\hline
\multirow{6}{*}{\begin{sideways}\emph{Per solved QBF}\end{sideways}} &
$\overline{a}$: & 667,906 & 596,840 & -10.6 \\
& $\overline{b}$: & 6,074 & 5,307 & -12.6 \\
& $\overline{t}$: & 4.75 & 2.81 & -40.8 \\
\cline{3-5}
 & $\tilde{a}$: & 827 & 828 & +0.1 \\
& $\tilde{b}$: & 1 & 1 & +0.0 \\
& $\tilde{t}$: & 1.31 & 0.37 & -71.7 \\
\hline
\end{tabular}
\end{minipage}
\caption{Average and median number of assignments ($\overline{a}$ and
$\tilde{a}$, respectively), backtracks ($\overline{b},\tilde{b}$), and 
workflow time ($\overline{t},\tilde{t}$) for planning instances from
Dungeon-\emph{v0} (left) and Dungeon-\emph{v1} (right) where both workflows using
DepQBF and incDepQBF found the optimal plan.}
\label{tab:compare:inc:noninc:dungeons}
\end{table}

Incremental solving performs particularly well on instances for which no plan exists. 
Considering the ten instances uniquely solved with incDepQBF
(Table~\ref{tab:compare:inc:noninc:dungeons:unique}), on average it took less
than 0.5 seconds to encode and solve one of the 201 unsatisfiable QBFs
(i.e., from plan length zero to the upper bound of 200) in the
 planning workflow. Considering the 13 instances solved using incDepQBF which
do not have a plan (Table~\ref{tab:compare:inc:noninc:dungeons:overview}), on
average the workflow took 73.80 seconds and incDepQBF spent 35,729
assignments and 135 backtracks. In contrast to that, the
workflow using DepQBF took 270.92 seconds on average to solve the five
instances which do not have a plan
(Table~\ref{tab:compare:inc:noninc:dungeons:overview}), and DepQBF
spent 421,619 assignments and 99 backtracks.

\subsection{Preprocessing}

The implementation of (inc)DepQBF does not include
preprocessing~\cite{DBLP:conf/cade/BiereLS11,DBLP:conf/sat/GiunchigliaMN10}. In
general, preprocessing
might be very beneficial for the performance of QBF-based workflows. The efficient
combination of preprocessing and incremental solving is part of ongoing
research in
SAT~\cite{DBLP:journals/fmsd/KupferschmidLSB11,DBLP:conf/sat/NadelRS14} and
QBF~\cite{DBLP:conf/cade/HeuleSB14,DBLP:conf/lpar/JanotaGM13,DBLP:conf/sat/MarinMB12,DBLP:conf/date/SeidlK14}.

In order to evaluate the \emph{potential} impact of preprocessing 
in our workflow, we carried out the following experiment. We ran the workflow
using DepQBF on all 288 planning instances with a time limit of 900
seconds and collected all QBFs that were generated this way. Like for the
results in Table~\ref{tab:compare:inc:noninc:dungeons:overview}, we ran
DepQBF and incDepQBF on these QBFs within our workflow. Additionally,
we ran the QBF solver Nenofex~\cite{DBLP:conf/sat/LonsingB08} because it performed well on QBFs generated from
the Dungeon benchmark.\footnote{\mbox{Results of Nenofex in the QBF Gallery 
2013:\ \ }\url{http://www.kr.tuwien.ac.at/events/qbfgallery2013/sc_apps/conf_planning_dungeon.html}}
Nenofex successively eliminates variables in a QBF by expansion at the cost of a possibly exponential
blow up of the formula size.
Figure~\ref{fig:times:inc:noninc:standalone} shows the run times of
DepQBF,  \mbox{incDepQBF}, Nenofex and DepQBF-pre, which combines 
DepQBF with the preprocessor
Bloqqer~\cite{DBLP:conf/cade/BiereLS11}. 
We accumulated the solving times spent on QBFs that were generated from a
particular planning instance. The plot shows these accumulated times for each
planning instance. Run times smaller than the time out of 900 seconds do not
necessarily indicate that the planning instance was solved because we
considered only a subset of the QBFs corresponding to the planning
instance. The performance of DepQBF-pre and Nenofex shown in
Figure~\ref{fig:times:inc:noninc:standalone} illustrates the  
benefits of preprocessing in the planning workflow. Among other techniques,
Bloqqer applies expansion, the core technique used in Nenofex, in a way that restricts the blow up of the formula size~\cite{DBLP:conf/sat/Biere04a,DBLP:conf/sat/BubeckB07}. 

 Given the results shown in Figure~\ref{fig:times:inc:noninc:standalone},
preprocessing might considerably improve the performance of incremental QBF
solving in our workflow. To this end, it is necessary to combine QBF
preprocessing and solving in an \emph{incremental} way.

\subsection{Comparison to Heuristic Approaches}

Although our focus is on a comparison of non-incremental and incremental QBF
solving, we report on additional experiments with the heuristic planning
tools ConformantFF~\cite{HoffmannB06} and T0~\cite{PalaciosG09}. In contrast to our implemented QBF-based approach
to conformant planning, heuristic tools do not guarantee to find a plan with the
optimal (i.e., shortest) length. In practical settings, plans with optimal
length are desirable. Moreover, the QBF-based approach allows to
 verify the non-existence of a plan with respect to an upper
bound on the plan length. Due to these differences, a comparison based on the
run times and numbers of solved instances only is not appropriate. 

Related to 
Table~\ref{tab:compare:inc:noninc:dungeons:overview}, ConformantFF solved 169
planning instances, where it found a plan for 144 instances and concluded that
no plan exists (with a length shorter than our considered upper bound of 200)
for 25 instances. Considering the 124 instances where both incDepQBF and
ConformantFF found a plan, for 42 instances the optimal plan found by
incDepQBF was strictly shorter than the plan found by ConformantFF. On the 124
instances, the average (median) length of the plan found by incDepQBF was 2.06
(1), compared to an average (median) length of 3.45 (1) by ConformantFF.

Due to technical problems, we were not able to run the experiments with
T0\footnote{Experiments with T0 were run on AMD Opteron 6176 SE, 2.3 GHz,
64-bit Linux} on the same system as the experiments with (inc)DepQBF and
ConformantFF. Hence the results by T0 reported in the following are actually
incomparable to Table~\ref{tab:compare:inc:noninc:dungeons:overview}. However,
we include them here to allow for a basic comparison of the plan lengths.

Using the same time and memory limits as for incDepQBF and ConformantFF, T0
solved 206 planning instances, where it found a plan for 203 instances and
concluded that no plan exists (with a length shorter than the upper bound of
200) for three instances.  Given the 156 instances where both incDepQBF
and T0 found a plan, for 56 instances the optimal plan found by incDepQBF was
strictly shorter than the plan found by T0. On the 156 instances, the average
(median) length of the plan found by \mbox{incDepQBF} was 2.25 (1), compared to an
average (median) length of 3.08 (2) by T0.

From the 13 instances solved by incDepQBF for which no plan exists
(Table~\ref{tab:compare:inc:noninc:dungeons:overview}), none was solved using
T0 and 12 were solved using ConformantFF.

Our experiments confirm that the QBF-based approach to conformant planning
finds optimal plans in contrast to the plans found by the heuristic
approaches implemented in ConformantFF and T0. 
 Moreover,  (inc)DepQBF and other search-based QBF solvers rely
on Q-resolution~\cite{DBLP:journals/iandc/BuningKF95} as the underlying proof
system. Given a Q-resolution proof $\Pi$ of the unsatisfiability of a QBF
$\psi$, it is possible to extract from $\Pi$ a
countermodel~\cite{DBLP:journals/fmsd/BalabanovJ12} or
strategy~\cite{DBLP:conf/ijcai/GoultiaevaGB11} of $\psi$ in terms of a set of
Herbrand functions. Intuitively, an Herbrand function
$f_{y}(x_{y_{1}},\ldots,x_{y_{n}})$ represents the values that a universal
variable $f_{y}$ must take to falsify $\psi$ with respect to the values of all
existential variables $x_{y_{1}},\ldots,x_{y_{n}}$ with $x_{y_{i}} < y$ in the
prefix ordering. Given a conformant planning problem $P$, Q-resolution proofs
and Herbrand function countermodels allow to independently explain and
verify~\cite{DBLP:conf/sat/NiemetzPLSB12} the non-existence of a plan (of a
particular length) for $P$ by verifying the unsatisfiability of the QBF
encoding of $P$. This is an appealing property of the QBF-based approach. In
practical applications, it may be interesting to have
an explanation of the non-existence of a plan in addition to the mere answer
that no plan exists.

The exact QBF-based approach for conformant planning can be combined with
heuristic approaches in a portfolio-style system, for example. Thereby, the two approaches
are applied in parallel and independently from each other. This way, modern
multi-core hardware can naturally be exploited.

\section{Conclusion}
\label{sect:conclusion}

We presented a case study of incremental QBF solving based on a workflow to
incrementally encode planning problems into sequences of QBFs. Thereby, we
focused on a general-purpose QBF solver. The incremental approach avoids some 
redundancy.  First, parts of the QBF encodings of shorter plan lengths can be
reused in the encodings of longer plan lengths. Second, the incremental QBF
solver benefits from information that was learned from previously solved
QBFs. Compared to heuristic approaches, the QBF-based approach 
 has the advantage that it always finds the shortest plan and
it allows to verify the non-existence of a plan by 
Q-resolution proofs.

Using variants of the solver DepQBF, incremental QBF solving outperforms non-incremental QBF solving in the planning 
workflow in terms of solved instances and statistics like
the number of backtracks, assignments, and run time. The results of our experimental study independently motivate the use
of incremental QBF solving in applications other than planning.  We
implemented the Java interface DepQBF4J to integrate the solver DepQBF in
our planning tool. This interface is extensible and can be combined with
arbitrary Java applications.

The experiments revealed that keeping learned information in incremental QBF
solving might be harmful if the heuristics of the solver are negatively
influenced. Our observations merit a closer look on these heuristics when used
in incremental solving. 
In general, the combination of preprocessing and incremental
solving~\cite{DBLP:conf/cade/HeuleSB14,DBLP:conf/lpar/JanotaGM13,DBLP:journals/fmsd/KupferschmidLSB11,DBLP:conf/sat/MarinMB12,DBLP:conf/date/MarinMLB12,DBLP:conf/sat/NadelRS14,DBLP:conf/date/SeidlK14}
could improve the performance of QBF-based workflows.


\begin{thebibliography}{10}
\providecommand{\url}[1]{\texttt{#1}}
\providecommand{\urlprefix}{URL }

\bibitem{DBLP:conf/sat/AudemardLS13}
Audemard, G., Lagniez, J.M., Simon, L.: {Improving Glucose for Incremental SAT
  Solving with Assumptions: Application to MUS Extraction}. In: J{\"a}rvisalo,
  M., Van~Gelder, A. (eds.) SAT. LNCS, vol. 7962, pp. 309--317. Springer (2013)

\bibitem{DBLP:journals/fmsd/BalabanovJ12}
Balabanov, V., Jiang, J.H.R.: {Unified QBF certification and its applications}.
  Formal Methods in System Design  41(1),  45--65 (2012)

\bibitem{DBLP:journals/ai/BaralKT00}
Baral, C., Kreinovich, V., Trejo, R.: {Computational Complexity of Planning and
  Approximate Planning in the Presence of Incompleteness}. Artificial
  Intelligence  122(1-2),  241--267 (2000)

\bibitem{DBLP:conf/sat/Biere04a}
Biere, A.: {Resolve and Expand}. In: Hoos, H.H., Mitchell, D.G. (eds.) SAT
  (Selected Papers). LNCS, vol. 3542, pp. 59--70. Springer (2004)

\bibitem{DBLP:conf/cade/BiereLS11}
Biere, A., Lonsing, F., Seidl, M.: {Blocked Clause Elimination for QBF}. In:
  Bj{\o}rner, N., Sofronie-Stokkermans, V. (eds.) CADE. LNCS, vol. 6803, pp.
  101--115. Springer (2011)

\bibitem{DBLP:journals/ai/BlumF97}
Blum, A., Furst, M.L.: {Fast Planning Through Planning Graph Analysis}.
  Artificial Intelligence  90(1-2),  281--300 (1997)

\bibitem{DBLP:conf/sat/BubeckB07}
Bubeck, U., {Kleine B{\"u}ning}, H.: {Bounded Universal Expansion for
  Preprocessing QBF}. In: Marques-Silva, J., Sakallah, K.A. (eds.) SAT. LNCS,
  vol. 4501, pp. 244--257. Springer (2007)

\bibitem{DBLP:journals/jar/CadoliSGG02}
Cadoli, M., Schaerf, M., Giovanardi, A., Giovanardi, M.: {An Algorithm to
  Evaluate Quantified Boolean Formulae and Its Experimental Evaluation}.
  Journal of Automated Reasoning  28(2),  101--142 (2002)

\bibitem{DPLL}
Davis, M., Logemann, G., Loveland, D.: A {M}achine {P}rogram for
  {T}heorem-proving. Communications of the ACM  5(7),  394--397 (1962)

\bibitem{DBLP:journals/entcs/EenS03}
E{\'e}n, N., S{\"o}rensson, N.: {Temporal Induction by Incremental SAT
  Solving}. Electronic Notes in Theoretical Computer Science  89(4),  543--560
  (2003)

\bibitem{DBLP:conf/aisc/EglyKLP14}
Egly, U., Kronegger, M., Lonsing, F., Pfandler, A.: {Conformant Planning as a
  Case Study of Incremental {QBF} Solving}. In: Aranda{-}Corral, G.A., Calmet,
  J., Mart{\'{\i}}n{-}Mateos, F.J. (eds.) AISC. LNCS, vol. 8884, pp. 120--131.
  Springer (2014)

\bibitem{Egly2016}
Egly, U., Kronegger, M., Lonsing, F., Pfandler, A.: Conformant planning as a
  case study of incremental {QBF} solving. Annals of Mathematics and Artificial
  Intelligence pp. 1--25 (2016),
  \url{http://dx.doi.org/10.1007/s10472-016-9501-2}

\bibitem{DBLP:journals/jair/GiunchigliaNT06}
Giunchiglia, E., Narizzano, M., Tacchella, A.: {Clause/Term Resolution and
  Learning in the Evaluation of Quantified Boolean Formulas}. Journal of
  Artificial Intelligence Research  26,  371--416 (2006)

\bibitem{DBLP:conf/sat/GiunchigliaMN10}
Giunchiglia, E., Marin, P., Narizzano, M.: {sQueezeBF: An Effective
  Preprocessor for QBFs Based on Equivalence Reasoning}. In: Strichman, O.,
  Szeider, S. (eds.) SAT. LNCS, vol. 6175, pp. 85--98. Springer (2010)

\bibitem{DBLP:conf/ijcai/GoultiaevaGB11}
Goultiaeva, A., Van~Gelder, A., Bacchus, F.: {A Uniform Approach for Generating
  Proofs and Strategies for Both True and False QBF Formulas}. In: Walsh, T.
  (ed.) IJCAI. pp. 546--553. AAAI Press (2011)

\bibitem{DBLP:conf/cade/HeuleSB14}
Heule, M., Seidl, M., Biere, A.: {A Unified Proof System for QBF
  Preprocessing}. In: Demri, S., Kapur, D., Weidenbach, C. (eds.) IJCAR. LNCS,
  vol. 8562, pp. 91--106. Springer (2014)

\bibitem{HoffmannB06}
Hoffmann, J., Brafman, R.I.: {Conformant planning via heuristic forward search:
  A new approach}. Artificial Intelligence  170(6-7),  507--541 (2006)

\bibitem{DBLP:conf/lpar/JanotaGM13}
Janota, M., Grigore, R., Marques{-}Silva, J.: {On QBF Proofs and
  Preprocessing}. In: McMillan, K.L., Middeldorp, A., Voronkov, A. (eds.)
  LPAR-19. LNCS, vol. 8312, pp. 473--489. Springer (2013)

\bibitem{DBLP:journals/iandc/BuningKF95}
{Kleine B{\"u}ning}, H., Karpinski, M., Fl{\"o}gel, A.: {Resolution for
  Quantified Boolean Formulas}. Information and Computation  117(1),  12--18
  (1995)

\bibitem{KroneggerPP13}
Kronegger, M., Pfandler, A., Pichler, R.: Conformant planning as a benchmark
  for {QBF}-solvers. In: Report Int. Workshop on Quantified Boolean Formulas
  (QBF 2013). pp. 1--5 (2013),
  \url{http://fmv.jku.at/qbf2013/reportQBFWS13.pdf}

\bibitem{DBLP:journals/fmsd/KupferschmidLSB11}
Kupferschmid, S., Lewis, M.D.T., Schubert, T., Becker, B.: {Incremental
  Preprocessing Methods for Use in BMC}. Formal Methods in System Design
  39(2),  185--204 (2011)

\bibitem{DBLP:conf/sat/LagniezB13}
Lagniez, J.M., Biere, A.: {Factoring Out Assumptions to Speed Up MUS
  Extraction}. In: J{\"a}rvisalo, M., Van~Gelder, A. (eds.) SAT. LNCS, vol.
  7962, pp. 276--292. Springer (2013)

\bibitem{DBLP:conf/tableaux/Letz02}
Letz, R.: {Lemma and Model Caching in Decision Procedures for Quantified
  Boolean Formulas}. In: Egly, U., Ferm{\"u}ller, C.G. (eds.) TABLEAUX. LNCS,
  vol. 2381, pp. 160--175. Springer (2002)

\bibitem{DBLP:conf/sat/LonsingB08}
Lonsing, F., Biere, A.: {Nenofex: Expanding NNF for QBF Solving}. In: {Kleine
  B{\"u}ning}, H., Zhao, X. (eds.) SAT. LNCS, vol. 4996, pp. 196--210. Springer
  (2008)

\bibitem{DBLP:conf/cp/LonsingE14}
Lonsing, F., Egly, U.: {Incremental QBF Solving}. In: O'Sullivan, B. (ed.) CP.
  LNCS, vol. 8656, pp. 514--530. Springer (2014)

\bibitem{DBLP:conf/icms/LonsingE14}
Lonsing, F., Egly, U.: {Incremental QBF Solving by DepQBF}. In: Hong, H., Yap,
  C. (eds.) ICMS. LNCS, vol. 8592, pp. 307--314. Springer (2014)

\bibitem{DBLP:conf/sat/LonsingEG13}
Lonsing, F., Egly, U., Van~Gelder, A.: {Efficient Clause Learning for
  Quantified Boolean Formulas via QBF Pseudo Unit Propagation}. In:
  J{\"a}rvisalo, M., Van~Gelder, A. (eds.) SAT. LNCS, vol. 7962, pp. 100--115.
  Springer (2013)

\bibitem{DBLP:conf/sat/MarinMB12}
Marin, P., Miller, C., Becker, B.: {Incremental QBF Preprocessing for Partial
  Design Verification - (Poster Presentation)}. In: Cimatti, A., Sebastiani, R.
  (eds.) SAT. LNCS, vol. 7317, pp. 473--474. Springer (2012)

\bibitem{DBLP:conf/date/MarinMLB12}
Marin, P., Miller, C., Lewis, M.D.T., Becker, B.: {Verification of Partial
  Designs using Incremental QBF Solving}. In: Rosenstiel, W., Thiele, L. (eds.)
  DATE. pp. 623--628. IEEE (2012)

\bibitem{DBLP:conf/sat/NadelRS14}
Nadel, A., Ryvchin, V., Strichman, O.: {Ultimately Incremental SAT}. In: Sinz,
  C., Egly, U. (eds.) SAT. LNCS, vol. 8561, pp. 206--218. Springer (2014)

\bibitem{DBLP:conf/sat/NiemetzPLSB12}
Niemetz, A., Preiner, M., Lonsing, F., Seidl, M., Biere, A.: {Resolution-Based
  Certificate Extraction for QBF - (Tool Presentation)}. In: Cimatti, A.,
  Sebastiani, R. (eds.) SAT. LNCS, vol. 7317, pp. 430--435. Springer (2012)

\bibitem{PalaciosG09}
Palacios, H., Geffner, H.: {Compiling Uncertainty Away in Conformant Planning
  Problems with Bounded Width}. Journal of Artificial Intelligence Research
  35,  623--675 (2009)

\bibitem{DBLP:conf/aaai/Rintanen07}
Rintanen, J.: {Asymptotically Optimal Encodings of Conformant Planning in
  {QBF}}. In: Holte, R.C., Howe, A.E. (eds.) AAAI. pp. 1045--1050. AAAI Press
  (2007)

\bibitem{DBLP:conf/date/SeidlK14}
Seidl, M., K{\"o}nighofer, R.: {Partial witnesses from preprocessed quantified
  Boolean formulas}. In: DATE. pp. 1--6. IEEE (2014)

\bibitem{SmithW98}
Smith, D.E., Weld, D.S.: {Conformant Graphplan}. In: Mostow, J., Rich, C.
  (eds.) AAAI/IAAI. pp. 889--896. {AAAI} Press / The {MIT} Press (1998)

\bibitem{DBLP:conf/cp/ZhangM02}
Zhang, L., Malik, S.: {Towards a Symmetric Treatment of Satisfaction and
  Conflicts in Quantified Boolean Formula Evaluation}. In: Hentenryck, P.V.
  (ed.) CP. LNCS, vol. 2470, pp. 200--215. Springer (2002)

\end{thebibliography}
\end{document}